\documentclass[lettersize,journal,compsoc]{IEEEtran}
\usepackage{amsmath,amsfonts}
\usepackage{algorithmic}
\usepackage{algorithm}
\usepackage{array}
\usepackage[caption=false,font=normalsize,labelfont=sf,textfont=sf]{subfig}
\usepackage{textcomp}
\usepackage{stfloats}
\usepackage{url}
\usepackage{verbatim}
\usepackage{graphicx}
\usepackage{cite}
\usepackage{caption}
\usepackage{tikz} \usetikzlibrary{quantikz}

\usepackage{braket}
\usepackage{enumitem}	% to use noindent like  \begin{itemize}[leftmargin=*] ... \end{itemize}
\setlist[itemize]{leftmargin=*}		% make the noindent change global

\usepackage[T1]{fontenc}	% fix bibtex error with things like: Rozp{\k{e}}dek

\usepackage{tikz}	
\usetikzlibrary{backgrounds,fit,decorations.pathreplacing}  % TikZ libraries
\usetikzlibrary{automata} % LATEX and plain TEX
\usetikzlibrary{circuits.logic.CDH}
\usetikzlibrary{circuits.logic.US}
\usetikzlibrary{mindmap}
\usetikzlibrary{decorations}
\usetikzlibrary{decorations.pathmorphing}

\usepackage{circuitikz}
\usetikzlibrary{arrows,decorations.pathreplacing}

\usepackage[colorinlistoftodos,prependcaption]{todonotes}
\usepackage{xcolor}
\usepackage{xargs}

\newcommandx{\yellownote}[2][1=]{\todo[inline,linecolor=yellow,backgroundcolor=yellow!25,bordercolor=yellow,#1]{#2}}
\newcommandx{\rednote}[2][1=]{\todo[inline,linecolor=red,backgroundcolor=red!25,bordercolor=red,#1]{#2}}
\newcommandx{\greennote}[2][1=]{\todo[inline,linecolor=green,backgroundcolor=green!25,bordercolor=green,#1]{#2}}

\begin{document}

 \title{Harnessing  Coding Theory for Reliable Network Quantum Communication}

\author{
Ching-Yi Lai,~\IEEEmembership{Senior Member,~IEEE},
and Kao-Yueh Kuo,~\IEEEmembership{Member,~IEEE}
}

\maketitle

\begin{abstract}

This article explores the application of coding techniques for fault-tolerant quantum computation and extends their usage to fault-tolerant quantum communication. We review repeater-based quantum networks, emphasizing the roles of coding theory and fault-tolerant quantum operations, particularly in the context of quantum teleportation. We highlight that fault-tolerant implementation of the Bell measurement enables reliable quantum communication without requiring a universal set of quantum gates. Finally, we discuss various quantum code candidates for achieving higher transmission rates.

\end{abstract}

% \begin{IEEEkeywords}
% Quantum communication, teleportation, quantum repeater,  quantum error correction, fault-tolerant quantum operations,
% \end{IEEEkeywords}

% \tableofcontents
 
\section{Introduction}

The field of quantum computing has revolutionized the computational landscape by offering unprecedented computing power compared to classical computers. The establishment of a quantum network, consisting of interconnected local quantum processors, opens up new horizons for applications beyond the realm of traditional Internet, including secure classical and quantum communications, delegated quantum computation, and distributed quantum computing~\cite{WER18}.

  \begin{figure}[h!]
    \centering	\includegraphics[width=0.45\textwidth]{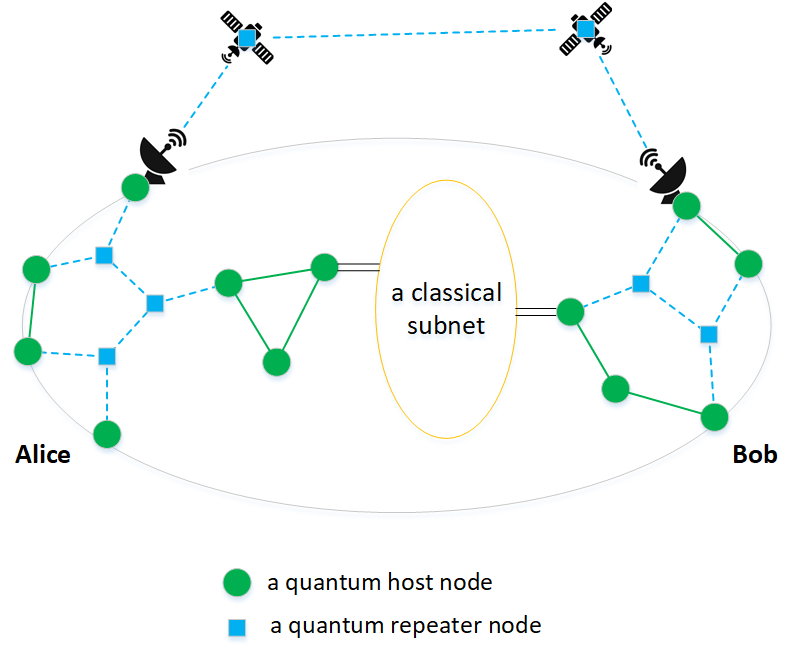}
    \caption{
        {Quantum Internet with quantum host nodes and repeater nodes.}
    } \label{fig:qnet}
    \end{figure}

 Figure~\ref{fig:qnet} depicts the concept of a future quantum internet, where multiple distributed hosts are interconnected through direct quantum channels, predominantly implemented using optical fibers.
  In situations where a direct quantum channel is limited by physical resources, several repeaters are employed to establish connections between distant host nodes. Each  host node possesses the capability to process general quantum information, while a repeater node specializes in specific quantum functions. 
Moreover, reliable quantum memories are assumed to be present in each node, ensuring the availability of quantum resources for communication tasks.
We assume that \textit{two-way} \textit{classical communication} is feasible between any two nodes in the network. This means that each node can transmit classical bits to any other node, utilizing channels such as the classical Internet. Additionally, quantum communication through satellites is also a viable means of connecting two distant quantum networks.

The quantum internet involves two types of quantum communication: the transmission of quantum information and quantum-assisted communication of classical information.
The transmission of quantum messages is a fundamental aspect that goes beyond the capabilities of the traditional Internet, enabling various advanced tasks. However, there are scenarios where high-rate classical communication is required via the quantum internet.
In this article we discuss how   to reliably implement these two types of quantum communication.

Quantum information is encoded and processed using delicate quantum states, which are highly vulnerable to disturbances.
We would like to transmit quantum information using such quantum states through lossy quantum channels. 
Moreover, local quantum operations are susceptible to imperfections. Nonetheless, we would like to have reliable quantum communication. 
To address these challenges, extensive research has focused on fault-tolerant quantum operations through quantum error-correcting codes~\cite{NC00}.

This article aims to review the concept of coding theory and elucidate how to  implement   a fault-tolerant quantum internet~\cite{Mur+14}. Each quantum function involved in the quantum internet, including \textit{quantum teleportation}, \textit{superdense coding},  and \textit{entanglement swapping}, can be integrated with quantum error correction techniques and will be explained in the following. Specifically, we focus on the utilization of teleportation-based error correction. Furthermore, we compare and analyze different quantum codes for achieving reliable teleportation-based quantum communication.
 Effective sparse-graph quantum codes, particularly when nonlocal qubit connectivity is accessible, have the potential to generate EPR pairs at notably high rates. These advancements present promising opportunities to enhance the efficiency of quantum communication systems.

\section{Network quantum communication}

A \textit{qubit} serves as the fundamental unit of quantum information. Similar to a classical bit, a qubit has two basis states denoted as $\ket{0}$ and $\ket{1}$. However,  qubits can also exist in a \textit{superposition} state, which is a linear combination of $\ket{0}$ and $\ket{1}$. When measuring a qubit, we obtain a one-bit outcome that provides information about its value. The probability distribution of this outcome is determined by the coefficients associated with the $\ket{0}$ and $\ket{1}$ basis states in the qubit's superposition.

\textit{Entanglement} is a notable quantum phenomenon that can occur between multiple qubits. One example of a two-qubit entangled state is the  Einstein--Podolsky--Rosen (EPR) pair, where there is an equal probability of both qubits being in the state $\ket{0}$ or both qubits being in the state $\ket{1}$. Remarkably, this entanglement remains regardless of the distance between the qubits. Thus EPR pairs shared between two distant parties allow  them to perform  protocols involving {local operations and classical communication (LOCC)} that are essential  for quantum communication.

  \begin{figure*}[htb!]
    \centering	\includegraphics[width=1.0\textwidth]{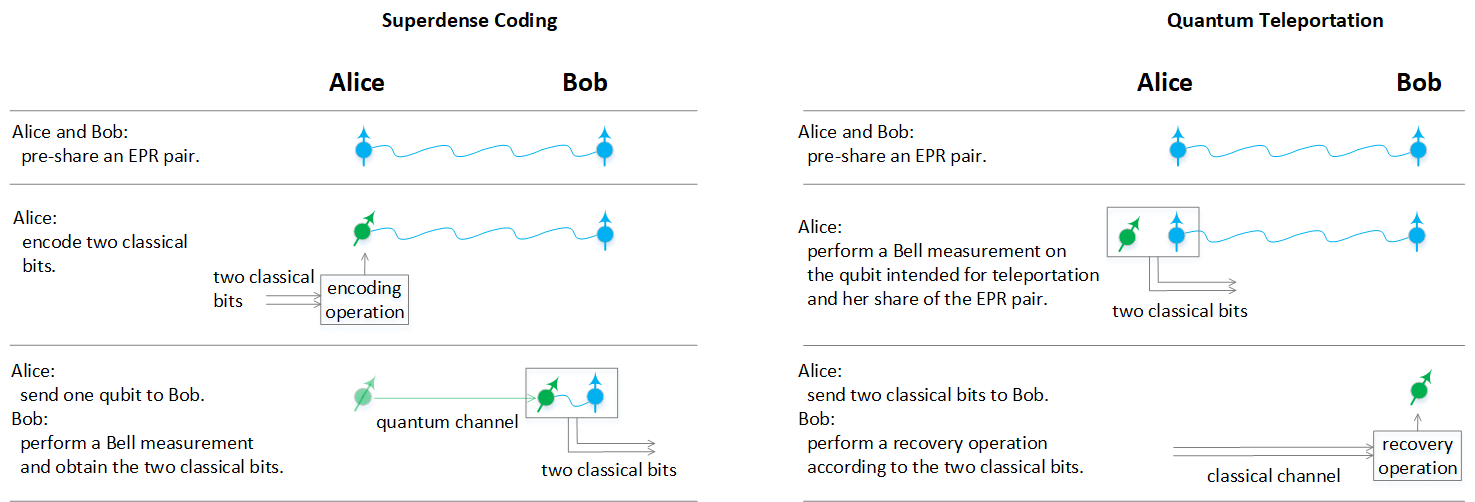}\\~\\
    (a)\qquad\qquad\qquad\qquad\qquad\qquad\qquad\qquad\qquad\qquad\qquad\qquad\qquad\qquad\qquad(b)
    \caption{
        (a)~Superdense coding. (b)~Quantum teleportation.  
    } \label{fig:tele_dense}
    \end{figure*}

Figure~\ref{fig:tele_dense}~(a) illustrates the concepts of superdense coding~\cite{BW92}, where two parties Alice and Bob share preexisting EPR pairs. By locally operating on  Alice's half of the EPR pair, it can be evolved into one of the four Bell states. Thus two classical bits are encoded in the two-qubit state. 
To be more precise, the encoding process applies quantum bit-flip and quantum phase-flip operations based on the values of the two bits. At this stage, Bob learns nothing if he measures his half of the EPR pair. Then, Alice sends her qubit to Bob through a qubit channel and Bob performs a Bell measurement to determine which Bell state on his hand. Overall, two classical bits are transmitted 
using one qubit channel and one pre-shared EPR pair.

Figure~\ref{fig:tele_dense}~(b) illustrates the concepts of   quantum teleportation~\cite{BBCJPW93}, where Alice and Bob once again share preexisting EPR pairs. In this scenario, Alice intends to transmit a qubit to Bob. Alongside her half of the EPR pair, she performs a Bell measurement on the two qubits. Consequently, Bob's half of the EPR pair is transformed into one of the four altered states corresponding to Alice's target qubit, but Bob remains unaware of the specific state until Alice discloses the two-bit measurement outcome. Subsequently, Bob applies a recovery operation, which involves quantum bit-flip and quantum phase-flip operations determined by the values of the two bits. In summary, this protocol enables the transmission of a qubit using two classical bit channels and one pre-shared EPR pair.

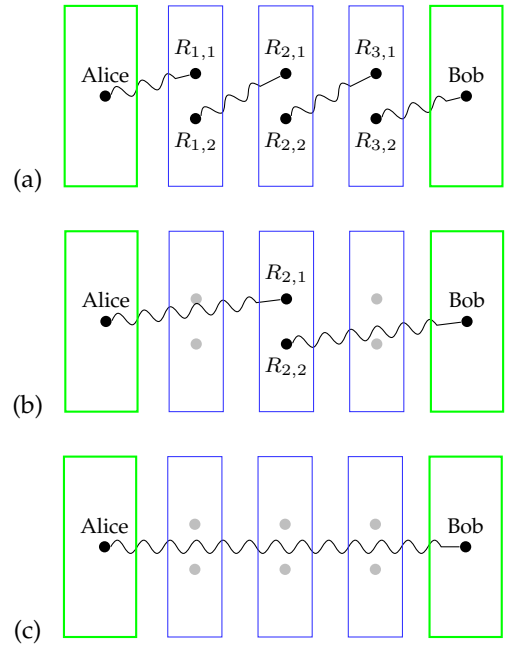
\begin{figure}[htbp]
	\centerline{ (a) ~
		\begin{tikzpicture}[scale=0.6][thick]
		\fontsize{8pt}{1} %%{font size}{line space}
		\tikzstyle{variablenode} = [draw,fill=white, shape=circle,minimum size=1.em]
		\node [fill=black,inner sep=1.5pt,shape=circle,label=above:Alice] (n0) at (-2,1.5) {} ;
		\node [fill=black,inner sep=1.5pt,shape=circle,label=above:Bob] (n1) at (6,1.5) {} ;
		\node [fill=black,inner sep=1.5pt,shape=circle,label=above:$R_{1,1}$] (n21) at (0,2) {} ;
 		\node [fill=black,inner sep=1.5pt,shape=circle,label=below:$R_{1,2}$] (n22) at (0,1) {} ;
 		\node [fill=black,inner sep=1.5pt,shape=circle,label=above:$R_{2,1}$] (n31) at (2,2) {} ;
 		\node [fill=black,inner sep=1.5pt,shape=circle,label=below:$R_{2,2}$] (n32) at (2,1) {} ;
 		% \node [fill=black,inner sep=1.5pt,shape=circle,label=above:$R_{3,1}$] (n41) at (4,2) {} ;
 		% \node [fill=black,inner sep=1.5pt,shape=circle,label=below:$R_{3,2}$] (n42) at (4,1) {} ;
% 		\node [fill=black,inner sep=1.5pt,shape=circle,label=above:$R_{5,1}$] (n51) at (7,2) {} ;
% 		\node [fill=black,inner sep=1.5pt,shape=circle,label=below:$R_{5,2}$] (n52) at (7,1) {} ;
 		% \node [fill=black,inner sep=1.5pt,shape=circle,label=above:$R_{n-1,1}$] (n61) at (6,2) {} ;
		% \node [fill=black,inner sep=1.5pt,shape=circle,label=below:$R_{n-1,2}$] (n62) at (6,1) {} ;
 		\node [fill=black,inner sep=1.5pt,shape=circle,label=above:$R_{3,1}$] (n71) at (4,2) {} ;
 	    \node [fill=black,inner sep=1.5pt,shape=circle,label=below:$R_{3,2}$] (n72) at (4,1) {} ;
 	    % \node [fill=lightgray,inner sep=1.2pt,shape=circle,label=below:$R_{1,0}$] (c1) at (0,0) {} ;
 	    % \node [fill=lightgray,inner sep=1.2pt,shape=circle,label=below:$R_{2,0}$] (c2) at (2,0) {} ;		
 	    % \node [fill=lightgray,inner sep=1.2pt,shape=circle,label=below:$R_{3,0}$] (c3) at (4,0) {} ;		
 	    % \node [fill=lightgray,inner sep=1.2pt,shape=circle,label=below:$R_{n-1,0}$] (c4) at (6,0) {} ;		
 	    % \node [fill=lightgray,inner sep=1.2pt,shape=circle,label=below:$R_{n,0}$] (c5) at (8,0) {} ;	
   	    % \node [fill=lightgray,inner sep=1.2pt,shape=circle,label=below:$A_{0}$] (c0) at (-2,0) {} ;
   	    % \node [fill=lightgray,inner sep=1.2pt,shape=circle,label=below:$B_{0}$] (c6) at (10,0) {} ;			
 		\draw[decorate,decoration={coil,aspect=0}] (n0) -- (n21);
 		\draw[decorate,decoration={coil,aspect=0}] (n22) -- (n31); 		 		\draw[decorate,decoration={coil,aspect=0}] (n32) -- (n71);
 		\draw[decorate,decoration={coil,aspect=0}] (n72) -- (n1);
 			\draw[green,thick]  (-2.9,-0.5) rectangle (-1.3,3.5);
 		\draw[green,thick]  (5.2,-0.5) rectangle (6.8,3.5);
 		\draw[blue!80]  (1.4,-0.5) rectangle (2.6,3.5);
 		\draw[blue!80]  (-0.6,-0.5) rectangle (0.6,3.5);
 		% \draw  (3.4,-1) rectangle (4.6,3);
 		\draw[blue!80]  (3.4,-0.5) rectangle (4.6,3.5);
 		% \draw[blue!80]  (7.2,-0.5) rectangle (8.8,3.5);
 		% \draw[lightgray,thick] (c1)--(c2)  (c4)--(c2) (c5)--(c4) (c0)--(c1) (c5)--(c6); 
 		% \node [fill=white,inner sep=1.5pt,shape=circle] (an4) at (4,1.5) {$\dots$} ;
		\end{tikzpicture}
	}
\vspace{0.5cm}
 	\centerline{ (b) ~
  \begin{tikzpicture}[scale=0.6][thick]
		\fontsize{8pt}{1} %%{font size}{line space}
		\tikzstyle{variablenode} = [draw,fill=white, shape=circle,minimum size=1.em]
		\node [fill=black,inner sep=1.5pt,shape=circle,label=above:Alice] (n0) at (-2,1.5) {} ;
		\node [fill=black,inner sep=1.5pt,shape=circle,label=above:Bob] (n1) at (6,1.5) {} ;
		\node [fill=lightgray,inner sep=1.5pt,shape=circle,label=above:$ $] (n21) at (0,2) {} ;
 		\node [fill=lightgray,inner sep=1.5pt,shape=circle,label=below:$ $] (n22) at (0,1) {} ;
 		\node [fill=black,inner sep=1.5pt,shape=circle,label=above:$R_{2,1}$] (n31) at (2,2) {} ;
 		\node [fill=black,inner sep=1.5pt,shape=circle,label=below:$R_{2,2}$] (n32) at (2,1) {} ;
 		% \node [fill=black,inner sep=1.5pt,shape=circle,label=above:$R_{3,1}$] (n41) at (4,2) {} ;
 		% \node [fill=black,inner sep=1.5pt,shape=circle,label=below:$R_{3,2}$] (n42) at (4,1) {} ;
% 		\node [fill=black,inner sep=1.5pt,shape=circle,label=above:$R_{5,1}$] (n51) at (7,2) {} ;
% 		\node [fill=black,inner sep=1.5pt,shape=circle,label=below:$R_{5,2}$] (n52) at (7,1) {} ;
 		% \node [fill=black,inner sep=1.5pt,shape=circle,label=above:$R_{n-1,1}$] (n61) at (6,2) {} ;
		% \node [fill=black,inner sep=1.5pt,shape=circle,label=below:$R_{n-1,2}$] (n62) at (6,1) {} ;
 		\node [fill=lightgray,inner sep=1.5pt,shape=circle,label=above:$ $] (n71) at (4,2) {} ;
 	    \node [fill=lightgray,inner sep=1.5pt,shape=circle,label=below:$ $] (n72) at (4,1) {} ;
 	    % \node [fill=lightgray,inner sep=1.2pt,shape=circle,label=below:$R_{1,0}$] (c1) at (0,0) {} ;
 	    % \node [fill=lightgray,inner sep=1.2pt,shape=circle,label=below:$R_{2,0}$] (c2) at (2,0) {} ;		
 	    % \node [fill=lightgray,inner sep=1.2pt,shape=circle,label=below:$R_{3,0}$] (c3) at (4,0) {} ;		
 	    % \node [fill=lightgray,inner sep=1.2pt,shape=circle,label=below:$R_{n-1,0}$] (c4) at (6,0) {} ;		
 	    % \node [fill=lightgray,inner sep=1.2pt,shape=circle,label=below:$R_{n,0}$] (c5) at (8,0) {} ;	
   	    % \node [fill=lightgray,inner sep=1.2pt,shape=circle,label=below:$A_{0}$] (c0) at (-2,0) {} ;
   	    % \node [fill=lightgray,inner sep=1.2pt,shape=circle,label=below:$B_{0}$] (c6) at (10,0) {} ;			
 		\draw[decorate,decoration={coil,aspect=0}] (n0) -- (n31);
 		% \draw[decorate,decoration={coil,aspect=0}] (n22) -- (n31); 		 		\draw[decorate,decoration={coil,aspect=0}] (n32) -- (n71);
 		\draw[decorate,decoration={coil,aspect=0}] (n32) -- (n1);
 			\draw[green,thick]  (-2.9,-0.5) rectangle (-1.3,3.5);
 		\draw[green,thick]  (5.2,-0.5) rectangle (6.8,3.5);
 		\draw[blue!80]  (1.4,-0.5) rectangle (2.6,3.5);
 		\draw[blue!80]  (-0.6,-0.5) rectangle (0.6,3.5);
 		% \draw  (3.4,-1) rectangle (4.6,3);
 		\draw[blue!80]  (3.4,-0.5) rectangle (4.6,3.5);
 		% \draw[blue!80]  (7.2,-0.5) rectangle (8.8,3.5);
 		% \draw[lightgray,thick] (c1)--(c2)  (c4)--(c2) (c5)--(c4) (c0)--(c1) (c5)--(c6); 
 		% \node [fill=white,inner sep=1.5pt,shape=circle] (an4) at (4,1.5) {$\dots$} ;
		\end{tikzpicture}
	}
 \vspace{0.5cm}
 	\centerline{ (c) ~
  \begin{tikzpicture}[scale=0.6][thick]
		\fontsize{8pt}{1} %%{font size}{line space}
		\tikzstyle{variablenode} = [draw,fill=white, shape=circle,minimum size=1.em]
		\node [fill=black,inner sep=1.5pt,shape=circle,label=above:Alice] (n0) at (-2,1.5) {} ;
		\node [fill=black,inner sep=1.5pt,shape=circle,label=above:Bob] (n1) at (6,1.5) {} ;
		\node [fill=lightgray,inner sep=1.5pt,shape=circle,label=above:$ $] (n21) at (0,2) {} ;
 		\node [fill=lightgray,inner sep=1.5pt,shape=circle,label=below:$ $] (n22) at (0,1) {} ;
 		\node [fill=lightgray,inner sep=1.5pt,shape=circle,label=above:$ $] (n31) at (2,2) {} ;
 		\node [fill=lightgray,inner sep=1.5pt,shape=circle,label=below:$ $] (n32) at (2,1) {} ;
 		% \node [fill=black,inner sep=1.5pt,shape=circle,label=above:$R_{3,1}$] (n41) at (4,2) {} ;
 		% \node [fill=black,inner sep=1.5pt,shape=circle,label=below:$R_{3,2}$] (n42) at (4,1) {} ;
% 		\node [fill=black,inner sep=1.5pt,shape=circle,label=above:$R_{5,1}$] (n51) at (7,2) {} ;
% 		\node [fill=black,inner sep=1.5pt,shape=circle,label=below:$R_{5,2}$] (n52) at (7,1) {} ;
 		% \node [fill=black,inner sep=1.5pt,shape=circle,label=above:$R_{n-1,1}$] (n61) at (6,2) {} ;
		% \node [fill=black,inner sep=1.5pt,shape=circle,label=below:$R_{n-1,2}$] (n62) at (6,1) {} ;
 		\node [fill=lightgray,inner sep=1.5pt,shape=circle,label=above:$ $] (n71) at (4,2) {} ;
 	    \node [fill=lightgray,inner sep=1.5pt,shape=circle,label=below:$ $] (n72) at (4,1) {} ;
 	    % \node [fill=lightgray,inner sep=1.2pt,shape=circle,label=below:$R_{1,0}$] (c1) at (0,0) {} ;
 	    % \node [fill=lightgray,inner sep=1.2pt,shape=circle,label=below:$R_{2,0}$] (c2) at (2,0) {} ;		
 	    % \node [fill=lightgray,inner sep=1.2pt,shape=circle,label=below:$R_{3,0}$] (c3) at (4,0) {} ;		
 	    % \node [fill=lightgray,inner sep=1.2pt,shape=circle,label=below:$R_{n-1,0}$] (c4) at (6,0) {} ;		
 	    % \node [fill=lightgray,inner sep=1.2pt,shape=circle,label=below:$R_{n,0}$] (c5) at (8,0) {} ;	
   	    % \node [fill=lightgray,inner sep=1.2pt,shape=circle,label=below:$A_{0}$] (c0) at (-2,0) {} ;
   	    % \node [fill=lightgray,inner sep=1.2pt,shape=circle,label=below:$B_{0}$] (c6) at (10,0) {} ;			
 		\draw[decorate,decoration={coil,aspect=0}] (n0) -- (n1);
 		% \draw[decorate,decoration={coil,aspect=0}] (n22) -- (n31); 		 		\draw[decorate,decoration={coil,aspect=0}] (n32) -- (n71);
 		% \draw[decorate,decoration={coil,aspect=0}] (n32) -- (n1);
 			\draw[green,thick]  (-2.9,-0.5) rectangle (-1.3,3.5);
 		\draw[green,thick]  (5.2,-0.5) rectangle (6.8,3.5);
 		\draw[blue!80]  (1.4,-0.5) rectangle (2.6,3.5);
 		\draw[blue!80]  (-0.6,-0.5) rectangle (0.6,3.5);
 		% \draw  (3.4,-1) rectangle (4.6,3);
 		\draw[blue!80]  (3.4,-0.5) rectangle (4.6,3.5);
 		% \draw[blue!80]  (7.2,-0.5) rectangle (8.8,3.5);
 		% \draw[lightgray,thick] (c1)--(c2)  (c4)--(c2) (c5)--(c4) (c0)--(c1) (c5)--(c6); 
 		% \node [fill=white,inner sep=1.5pt,shape=circle] (an4) at (4,1.5) {$\dots$} ;
		\end{tikzpicture}
	}
 \caption{Repeater-based quantum channel.} \label{fig:repeater-based}
	\end{figure}

 To establish shared EPR pairs, one party generates local EPR pairs and transmits half of them to the other party via quantum channels. Subsequently, both parties engage in an LOCC  \textit{entanglement purification} protocol to extract a small number of clean EPR pairs. These EPR pairs need to be consistently prepared and stored in quantum memory, ensuring their availability for future applications.

  Long-distance quantum communication presents challenges due to noise interference and energy dissipation in quantum channels, which can lead to the destruction of quantum information. Quantum repeaters, similar to classical relays, are typically employed to enhance long-distance communication. However, 
  an arbitrary unknown quantum state cannot be cloned~\cite{NC00}, limiting the reproduction and amplification of certain quantum states. Fortunately, for quantum communication, the preparation of   EPR pairs is sufficient, regardless of the messages to be transmitted. Through the entanglement swapping technique, it is possible to create long-distance EPR pairs for communication, rendering the quantum internet repeater-based.

 Figure~\ref{fig:repeater-based} illustrates the concept of a repeater-based quantum channel, which enables the creation of a long-distance EPR pair between Alice and Bob. The process involves generating EPR pairs between adjacent nodes along the path from Alice to Bob.

Let us consider Repeater 1, which shares an EPR pair with Alice and another EPR pair with Repeater 2. By utilizing the principle of quantum teleportation, Repeater 1 can teleport one half of the EPR pair shared with Alice to Repeater 2, resulting in the creation of a shared EPR pair between Alice and Repeater 2. This procedure is known as an \textit{entanglement swapping}. Consequently, through a series of entanglement swappings facilitated by transmitting classical bits from Alice to Bob, a long-distance EPR pair can be established between them.

\section{Coding theory}

We briefly introduce the coding theory. In digital systems, information is typically represented using a string of bits. When transmitting a data string through a communication channel, there is a probability that each bit may be flipped due to the presence of noise and imperfections in the channel.

An error-correcting code is an encoding algorithm that adds redundancy to digital information to protect it from noise. By introducing additional bits, the code allows for the detection or correction of a certain number of errors. While it is not possible to eliminate all errors, an effective error-correcting code reduces the error rate below the original physical error rate, making it valuable in practical applications.

In general, a code is a collection of bit strings, called \textit{codewords}, which satisfy specific \textit{parity check} conditions. A parity check involves verifying that the number of ones in a certain subset of a codeword is even. If a bit string passes all the parity checks, it is considered a valid codeword. Conversely, for invalid words, the outcomes of parity checks provide meaningful information about the error operation, called \textit{error syndrome}. Based on the error syndrome, a decoding algorithm is employed to determine the most probable error and its corresponding recovery operation.

The minimum distance of a code represents the smallest number of errors that can transform a codeword into another codeword, which characterizes the code's error correction capability. 
Meanwhile, the code rate is the ratio of the number of information bits to the code length. 
An ideal code exhibits both a high rate and a high minimum distance.

One of the simplest error-correcting codes is the three-bit repetition (3-rep) code, which encodes 0 as 000 and 1 as 111.  The valid  codewords in the 3-rep code are 000 and 111, while 001, 010, 100, 011, 101, and 110 are considered invalid. It is evident that the parity of any two bits in a codeword is always even.
Using a majority vote, the 3-rep code can be decoded, enabling the correction of any single-bit error.
This code has rate $1/3$ and minimum distance $3$.

We next explain how the concepts of coding theory extended for quantum error correction~\cite{NC00}. 
In the quest to protect quantum information from errors, traditional coding techniques face obstacles due to the continuous nature of quantum noises and the limitations imposed by the no-cloning theorem. 
Fortunately,   orthogonal basis states can be cloned using elementary quantum gates. Additionally, by correcting a discrete set of quantum errors, it becomes feasible to correct their linear combinations.  This enables the development of coding theory specifically tailored to address quantum bit-flip and phase-flip errors.

Among the various quantum codes, Calderbank--Shor--Steane (CSS) stabilizer codes are particularly valuable. They serve as the quantum counterpart to the parity-check codes. In these codes, information qubits are combined with ancillary qubits to form quantum codewords that satisfy certain bit parity-checks and phase parity-checks. Notably, these parity checks can be performed through stabilizer measurements on the quantum state without disturbing its encoded information. The measurement outcomes provide the error syndrome, which can be utilized to deduce the most likely error.

An example of a simple CSS stabilizer code is the 9-qubit Shor code, which employs two instances of the 3-rep code to independently check for bit-flips and phase-flips~\cite{NC00}.

 A  quantum code can be defined by a graph with parity-check nodes and qubit variable nodes.
 Each parity-check node   examines the parity of qubit variable nodes which it is  connected to. 
When the number of edges in the graph is relatively small,   the resultant quantum code is termed a sparse-graph or low-density parity-check (LDPC) quantum code.
 
 A two-dimensional (2D) code structure is typically defined on a 2D lattice~\cite{DKLP02}, while the framework for 3D/4D codes follows a similar approach. In contrast, a general sparse-graph quantum code may not inherently possess such a defined topology.
 This distinction necessitates nonlocal interactions for its implementation.
 However, sparse-graph  quantum codes possess the potential for high code rates and considerable code distance.

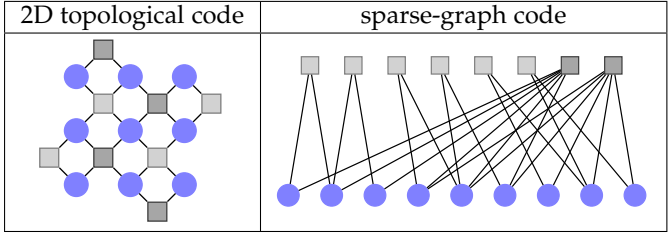
\begin{figure}%[htpb]
\centering 
\begin{tabular}{ |c|c| } 
\hline
2D topological code						& sparse-graph code		\\
\hline
\resizebox{0.1\textheight}{!}{\begin{tikzpicture}[node distance=1.3cm,>=stealth,bend angle=45,auto]

%\tikzstyle{chk}=[rectangle,draw=black!75,fill=black!75,minimum size=4mm]
%\tikzstyle{var}=[circle,   draw=black!75,fill=black!75,minimum size=4mm]

\tikzstyle{chk}=[rectangle,thick,draw=black!75,fill=black!35,minimum size=4mm]
\tikzstyle{chz}=[rectangle,thick,draw=black!50,fill=gray!35,minimum size=4mm]
\tikzstyle{var}=[circle,thick,draw=blue!50,fill=blue!50,minimum size=5mm,font=\footnotesize]

\tikzstyle{VAR}=[circle,thick,draw=blue!75,fill=blue!20,minimum size=5mm]
\tikzstyle{fac}=[anchor=west,font=\footnotesize]

\node[var] (x1) at (0 *4/7 ,  4 *4/7) {};
\node[var] (x2) at (2 *4/7 ,  4 *4/7) {};
\node[var] (x3) at (4 *4/7 ,  4 *4/7) {};
\node[var] (x4) at (0 *4/7 ,  2 *4/7) {};
\node[var] (x5) at (2 *4/7 ,  2 *4/7) {};
\node[var] (x6) at (4 *4/7 ,  2 *4/7) {};
\node[var] (x7) at (0 *4/7 ,  0 *4/7) {};
\node[var] (x8) at (2 *4/7 ,  0 *4/7) {};
\node[var] (x9) at (4 *4/7 ,  0 *4/7) {};
\node[chk] (c1) at (1 *4/7 ,  5 *4/7) {};
\node[chz] (c2) at (1 *4/7 ,  3 *4/7) {};
\node[chk] (c3) at (3 *4/7 ,  3 *4/7) {};
\node[chz] (c4) at (5 *4/7 ,  3 *4/7) {};
\node[chz] (c5) at(-1 *4/7 ,  1 *4/7) {};
\node[chk] (c6) at (1 *4/7 ,  1 *4/7) {};
\node[chz] (c7) at (3 *4/7 ,  1 *4/7) {};
\node[chk] (c8) at (3 *4/7 , -1 *4/7) {};

\node[fac] (xf) at (0 *4/7 ,  5.4 *4/7) {};	% add some white space at top

\draw[thick] (x1) -- (c1);
\draw[thick] (x2) -- (c1);
\draw[thick] (x1) -- (c2);
\draw[thick] (x2) -- (c2);
\draw[thick] (x4) -- (c2);
\draw[thick] (x5) -- (c2);
\draw[thick] (x2) -- (c3);
\draw[thick] (x3) -- (c3);
\draw[thick] (x5) -- (c3);
\draw[thick] (x6) -- (c3);
\draw[thick] (x3) -- (c4);
\draw[thick] (x6) -- (c4);
\draw[thick] (x4) -- (c5);
\draw[thick] (x7) -- (c5);
\draw[thick] (x4) -- (c6);
\draw[thick] (x5) -- (c6);
\draw[thick] (x7) -- (c6);
\draw[thick] (x8) -- (c6);
\draw[thick] (x5) -- (c7);
\draw[thick] (x6) -- (c7);
\draw[thick] (x8) -- (c7);
\draw[thick] (x9) -- (c7);
\draw[thick] (x8) -- (c8);
\draw[thick] (x9) -- (c8);

\end{tikzpicture}}	& \resizebox{0.2\textheight}{!}{\begin{tikzpicture}[node distance=1.3cm,>=stealth,bend angle=45,auto]

\tikzstyle{chk}=[rectangle,thick,draw=black!75,fill=black!35,minimum size=4mm]
\tikzstyle{chz}=[rectangle,thick,draw=black!50,fill=gray!35,minimum size=4mm]
\tikzstyle{var}=[circle,thick,draw=blue!50,fill=blue!50,minimum size=5mm,font=\footnotesize]

%\tikzstyle{chk}=[rectangle,thick,draw=black!75,fill=black!20,minimum size=4mm]
%\tikzstyle{var}=[circle,thick,draw=blue!75,fill=gray!20,minimum size=4mm,font=\footnotesize]
\tikzstyle{VAR}=[circle,thick,draw=blue!75,fill=blue!20,minimum size=5mm,font=\footnotesize]
\tikzstyle{fac}=[anchor=west,font=\footnotesize]

\node[var] (x1)  at (1-1, 0) {};
\node[var] (x2)  at (2-1, 0) {};
\node[var] (x3)  at (3-1, 0) {};
\node[var] (x4)  at (4-1, 0) {};
\node[var] (x5)  at (5-1, 0) {};
\node[var] (x6)  at (6-1, 0) {};
\node[var] (x7)  at (7-1, 0) {};
\node[var] (x8)  at (8-1, 0) {};
\node[var] (x9)  at (9-1, 0) {};

\node[chz] (c1) at  (1-1+0.5, 3) {};
\node[chz] (c2) at  (2-1+0.5, 3) {};
\node[chz] (c3) at  (3-1+0.5, 3) {};
\node[chz] (c4) at  (4-1+0.5, 3) {};
\node[chz] (c5) at  (5-1+0.5, 3) {};
\node[chz] (c6) at  (6-1+0.5, 3) {};
\node[chk] (c7) at  (7-1+0.5, 3) {};
\node[chk] (c8) at  (8-1+0.5, 3) {};

\node[fac] (xf) at  (0, -0.5) {};	% add some white space at bottom

\draw[thick] (x1)  -- (c1);
\draw[thick] (x2)  -- (c1);
\draw[thick] (x2)  -- (c2);
\draw[thick] (x3)  -- (c2);

\draw[thick] (x4)  -- (c3);
\draw[thick] (x5)  -- (c3);
\draw[thick] (x5)  -- (c4);
\draw[thick] (x6)  -- (c4);

\draw[thick] (x7)  -- (c5);
\draw[thick] (x8)  -- (c5);
\draw[thick] (x8)  -- (c6);
\draw[thick] (x9)  -- (c6);

\draw[thick] (x1)  -- (c7);
\draw[thick] (x2)  -- (c7);
\draw[thick] (x3)  -- (c7);
\draw[thick] (x4)  -- (c7);
\draw[thick] (x5)  -- (c7);
\draw[thick] (x6)  -- (c7);

\draw[thick] (x4)  -- (c8);
\draw[thick] (x5)  -- (c8);
\draw[thick] (x6)  -- (c8);
\draw[thick] (x7)  -- (c8);
\draw[thick] (x8)  -- (c8);
\draw[thick] (x9)  -- (c8);

\end{tikzpicture}}	\\
%\resizebox{0.075\textheight}{!}{\input{Fig_913_surface.tex}} & \resizebox{0.2\textheight}{!}{\input{Fig_15x20_Gallager.tex}}	\\
\hline
\end{tabular}
%%%%%%%%
% \footnotesize
% \begin{tabular}{ |c|c| } 
% \hline
% 2D topological code						& sparse-graph code		\\
% \hline
% suitable for qubits with locality       & require non local qubits \\
% \hline
% code rate vanishes                      & support constant code rate \\
% \hline
% \resizebox{0.08\textheight}{!}{\input{Fig_913_surface.tex}}	& \resizebox{0.16\textheight}{!}{\input{Fig_913_Shor.tex}}	\\
% \hline
% \end{tabular}
%%%%%%%%
    \caption{Examples of code structures, with circles representing qubit nodes and squares representing check nodes.}
    \label{fig:codes}
\end{figure}

To illustrate, Fig.~\ref{fig:codes} provides  examples of both a nine-qubit 2D surface code and the nine-qubit Shor code. In this representation, a circle represents a data qubit, a dark gray box represents a bit-flip check node, and a light gray box represents a phase-flip check node. Interactions between a qubit and a check node are depicted by connecting edges.

When devising a quantum code,  two crucial aspects come into play: Firstly, a code should allow feasible syndrome measurements. These measurements ideally involve only a small subset of qubits within a parity check. Sparse qubit interactions are preferred, as   multiple-qubit interactions can be challenging to implement. Furthermore, parity checks involving qubits in close proximity are more favorable as they are easier to implement.
Secondly, an efficient decoder is essential. Its decoding time should ideally be polynomial (preferably linear) in the code size. This becomes crucial as errors accumulate at the speed of gate operating times.

 There are many decoding algorithms.
For decoding of topological codes, the minimum-weight perfect matching (MWPM) algorithm has been successfully applied \cite{DKLP02}. In the case of  sparse-graph quantum codes, belief propagation (BP) decoding is commonly employed  due to its nearly linear complexity \cite{KL21}. Notably, BP decoding can be extended to correct a combination of data and measurement errors \cite{ALB20}. Moreover, this approach shows promise in addressing other error types, such as gate errors.

\section{Fault-tolerant quantum operations}

When Alice and Bob aim to employ encoded qubit communication, Alice needs to utilize an encoding circuit that involves single-qubit and two-qubit gates, state preparations, and possibly qubit measurements. However, in practice, all of these components are susceptible to faults, making it challenging to create the target qubit in the desired encoded state.

Upon receiving a channel-corrupted state, Bob is required to execute a decoding circuit and perform recovery operations. Nevertheless, the syndrome measurement outcomes will be affected by measurement imperfections. Moreover, the decoding quantum circuit itself, used for syndrome extraction, is subject to noise.

To achieve reliable quantum communication, the realization of fault-tolerant quantum operations, particularly the error correction block, becomes necessary. The fundamental concept is to replace each quantum operation with its encoded counterpart, and quantum error correction is continually applied to mitigate errors.
 Since errors may propagate through two-qubit gates, each encoded operation has to be carefully designed  to be \textit{fault-tolerant} so that  
 a small amount of errors will not evolve into an uncorrectable error for the underlying quantum error-correcting code.

\begin{figure}[h!]
%%\centering 
\resizebox{0.47\textwidth}{!}{\input{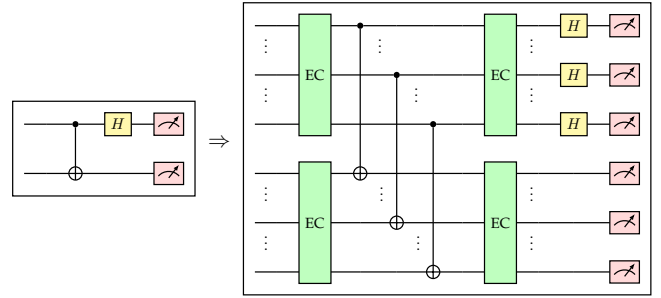}}
\caption{
     (left) Bell measurement. (right) Fault-tolerant implementation of the Bell measurement based on the nine-qubit Shor code. %The fault-tolerant implementation uses {bitwise single the transversal operation}. 
     Each EC block denotes an error correction for the nine-qubit code.
     % \knote{Are the blue words typos?}
}\label{fig:Ft_cnot}
\end{figure}

Figure~\ref{fig:Ft_cnot} illustrates the circuit implementation of the Bell measurement and its corresponding encoded version. In each circuit, qubits are represented by horizontal wires, and the gates are applied from left to right.
There are three types of gates: two-qubit controlled-NOT (CNOT) gates, Hadamard (H) gates, and single-qubit measurements.

In this particular example, a single qubit is encoded into nine physical qubits using the 9-qubit Shor code. As a result, the encoded version of the CNOT, Hadamard gates, and qubit measurements are implemented in a bitwise fashion. Note that any single component error will not propagate  within the same codeword, highlighting the fault-tolerant nature of this encoded circuit. The bitwise style of implementation is particularly significant for CSS stabilizer codes with equal bit-flip and phase-flip checks.

The error correction blocks (ECs) shown in Fig.~\ref{fig:Ft_cnot} consist of syndrome measurements, syndrome-based decodings, and recovery operations.  There are various methods for performing syndrome measurements. In this context, we will focus on the Knill syndrome extraction method~\cite{Knill05}, which is based on quantum teleportation in Fig.~\ref{fig:tele_dense} (b).

Suppose that Alice possesses an $n$-qubit encoded state and shares $n$ EPR pairs with Bob. In this scenario, Alice can utilize teleportation to transmit her $n$-qubit state to Bob. To complete the teleportation process, Bob needs to deduce a logical recovery operation based on Alice's Bell measurement outcomes.
Remarkably, these outcomes encapsulate the  information regarding the logical bit-flip and phase-flip operations required for teleportation. Furthermore, the error syndromes pertaining to the errors present in the transmitted encoded state are also incorporated into the Bell measurement outcomes. As a result, Bob can execute a logical recovery operation and perform error correction based on the information obtained from the Bell measurements.

Moreover, this  teleportation-based error correction technique  can be locally implemented, provided that a local node can generate a sufficient number of EPR pairs. The process relies solely on shared EPRs and Bell measurements, as shown in Fig.~\ref{fig:Ft_cnot}, which can be fault-tolerantly implemented (without the ECs). Given that each host or repeater node within the quantum internet needs to consistently generate numerous EPR pairs, its quantum memory can be effectively maintained through the utilization of this teleportation-based error correction mechanism.

\section{Reliable teleportation-based quantum communication}

In long-distance quantum communication or repeater-based quantum networks, repeater nodes play a crucial role in generating and distributing entangled pairs of qubits, as well as performing entanglement swapping and teleportation. 

To ensure the high quality of the shared entangled pairs, entanglement purification techniques are employed.
The  process begins with one node generating a large number of EPR pairs and sending half   to the other node. Both nodes independently perform bit-flip and phase-flip checks on their respective qubits, and exchange the outcomes of these checks. By analyzing the outcomes, they   identify and recover errors, thereby purifying a subset of the shared EPR pairs. This procedure resembles the decoding process of an error-correcting code, but it requires two-way classical communication between the nodes. Consequently, the rate of EPR generation is primarily limited by the delay in exchanging classical messages.

Entanglement purification utilizes basic quantum gates and two-qubit EPR states. However, it demands significant resources due to its continuous requirement.
For more efficient and reliable quantum communication, encoded EPR pairs are preferred. This allows for active quantum error correction to be executed while utilizing fewer resources.

An advanced approach assumes  that the host nodes possess the capability to encode and decode specific CSS stabilizer codes.    Two adjacent nodes first generate   high-quality EPR pairs using entanglement purification with two-way classical communication. Subsequently, one node locally encodes an EPR pair, and teleports half of it to the other node using one-way classical communication. 
Errors in the encoded EPR are corrected locally, allowing the creation of encoded entangled pairs between adjacent nodes.  
This method eliminates the need for two-way classical communication during quantum communication, resulting in a higher communication rate~\cite{JTN+09}.

Another advanced approach involves direct quantum communication, assuming robust quantum computation capabilities with large CSS stabilizer codes and reliable quantum channels~\cite{Fow+10}. Here, one host node directly transmits encoded qubits through quantum channels, and the receiver node employs fault-tolerant teleportation-based error correction to recover the state. This process bypasses the need for prior entanglement purification and effectively handles channel losses through the quantum code.

Both advanced approaches enable the creation of an encoded version of the repeater-based quantum channel  in Fig.~\ref{fig:repeater-based}, facilitating fault-tolerant quantum communication. However,  their effectiveness relies on maintaining a physical error rate lower than the error threshold of the underlying CSS stabilizer code. 
Consequently, the hardware requirements for achieving fault-tolerant quantum communication using CSS stabilizer codes remain quite demanding in current technology. The need to maintain a low physical error rate poses significant challenges for practical implementation.

\section{Reliable quantum-assisted classical communication}

  Similarly to the previous discussion, the central component of superdense coding is the Bell measurement, which can be fault-tolerantly implemented using a CSS stabilizer code. In the scenario where Alice and Bob share an EPR pair, Alice encodes her half of the pair into a CSS code and performs logical bit-flip and phase-flip operations based on the two bits to be transmitted. Bob then performs a half-logical Bell measurement on his half of the EPR pair and the received qubits to decode the classical bits.

This scenario falls under the realm of entanglement-assisted quantum coding theory~\cite{BDH06}, focusing on the transmission of classical bits.  It can be assumed that the error rate on Bob's halves of the EPR pairs can be lower due to quantum memory protection, using teleportation-based error correction with local EPR pairs.

While all the components in this setting can be fault-tolerantly implemented, a thorough discussion on resource tradeoff analysis is necessary to ascertain the quantum advantage of a fault-tolerant implementation of quantum-assisted classical communication. Without such analysis, the benefits might not justify the costs, particularly in instances demanding ultra-high rate communication.

\section{Candidates of quantum codes for practical implementation}

In fault-tolerant quantum computing, it is important to consider various error sources during the error-correction procedure. These include  data errors, gate errors, ancilla preparation errors, and measurement errors. Accounting for these errors in every error syndrome extraction round establishes the circuit noise model. 
Moreover, the practical complexity of decoding a quantum code is contingent on the fault-tolerant error correction procedure. 
Consequently, quantum codes that enable simple fault-tolerant error correction implementations and efficient decoding are highly desirable.

Many near-term quantum systems exhibit high-fidelity quantum operations primarily for locally-connected qubits. 
For instance, in superconducting qubits, interaction between nearby qubits is feasible, while establishing connections between distant qubits presents considerable challenges. Trapped-ion qubits or neutral-atom qubits permit some degree of nonlocal interaction. However, establishing long-distance interactions between qubits often incurs high costs, leading to inevitable fidelity loss.
Photonic qubits, also known as flying qubits, possess the potential for long-distance entanglement generation and transmission due to their inherent mobility. Yet, establishing connectivity between distant flying qubits is more challenging and may necessitate specialized techniques for efficient long-range interactions.

In such scenarios,  2D topological codes offer suitable solutions. These codes can be realized on a planar layout, necessitating minimal qubit interactions. Surface codes, a prominent example, exhibit a rectangular grid structure and employ small parity checks on four qubits. This design choice balances gate density and expedites syndrome extraction. However, their code rate diminishes as the code length increases.

In future quantum systems with enhanced qubit connectivity, the adoption of higher-dimensional topological codes or more generalized sparse-graph quantum codes becomes feasible. These advancements unlock the potential for \textit{single-shot} decoding, enabling effective error correction through every round of noisy syndrome extraction. This immediate error treatment prevents severe error accumulation and simplifies error analysis.

Specifically, the aforementioned teleportation-based error correction allows for single-shot decoding~\cite{ZLBK20}. It relies on an equivalent number of EPR pairs as the code size, and the outcomes of Bell measurements constitute all the error syndromes. Thus, the decoding complexity for this single-shot error-correction method remains within the same order of complexity.

Conversely, when each parity check is individually measured (utilizing raw syndrome extraction or Shor-type syndrome extraction~\cite{NC00}), multiple rounds of noisy syndrome measurements, typically proportional to the minimum distance, are necessary to obtain reliable error syndromes. 
This effectively increases the number of error variables, which significantly amplifies the decoding complexity. However, this  can be circumvented if certain nonlocal interactions are allowed, facilitating teleportation-based quantum error correction.

Numerous technologies assume local connections as the foundation for establishing accurate quantum memories and facilitating nonlocal logical connections. Within this framework, topological codes serve as the foundational correction mechanism, upon which sparse-graph quantum codes are implemented. This setup can be considered as a concatenated coding system. Exploring cross-layer optimization for overhead and performance is a future research focus.

\subsection{EPR generation rate}

Utilizing high code-rate quantum codes  presents notable advantages in boosting  EPR generation rates, particularly in large-scale quantum systems featuring arbitrary qubit connectivity.
To illustrate, let us consider the EPR generation rate when a host node receives half of the logical EPR pairs, protected by a specific quantum code, and uses teleportation-based error correction to rectify these pairs. For simplicity, we will exclude the overhead involved in encoding EPR states at the source node.

The rate of EPR generation at the receiving node hinges on the speed of error correction, where local EPR pairs are initially generated, followed by Bell measurements on halves of these local pairs and the received qubits. A round of error correction, comprising an ancilla preparation step, two CNOT steps, and one measurement step, typically takes approximately $4T$ time units. Here, $T$ represents the time unit for elementary gate operations or qubit measurements determined by the underlying quantum technology.

In this model, errors occurring in gates and measurements can  translate into data errors, leading to a higher effective  data error rate~\cite{ZLBK20}. We perform a rough estimation of this effective data error rate, assuming all gates and measurements have the same error rate $p_g$, while the EPR communication error rate is $p_c$. In teleportation-based error correction, the outcome of a single-qubit measurement in the Bell measurement is impacted by two ancillary state preparations, two CNOT gates, one single-qubit measurement, and the communication error in the EPR pair. A pessimistic estimate of the effective error rate could be $p_c+5p_g$.

We compare the EPR generation rates between topological codes and sparse-graph codes. Assuming communication errors are dominant with an error rate of $p_c=5\%$, while gates have a less severe error rate of $p_g=0.1\%$, the effective data error rate roughly amounts to $5.5\%$. Our target is to achieve a decoding logical error rate of around $10^{-4}$.

 Let us consider a scenario where the receiving node has a large-scale quantum device capable of processing 68,200 physical qubits. In teleportation-based error correction of an $n$-qubit code, $2n$ ancillary qubits are required. Thus the host node can process roughly $68200/3n$ code blocks.  

First, consider the 2D rotated surface codes, where one information qubit is encoded in a code block. To meet our criteria, a surface code with  minimum distance   $d=17$ is needed~\cite[Fig.~5]{KL21}, and its code length is $n=d^2=289$. Consequently, the host node can simultaneously process approximately $78$ code blocks. Decoding this many blocks every $4T$ yields an EPR generation rate of $78/4 = 19.5$ units per $T$ time.

Next, let us examine a random sparse-graph code with a length of $n=3786$ and a rate of $1/4$, which meets our performance criteria~\cite[Fig.~2]{KL21}. 
This code has low density of parity checks, where each parity-check operates on an average of 16 qubits. The receiving node can process roughly six code blocks. However, each code block encodes 946 information qubits. Therefore, the EPR generation rate becomes $946 \times 6 / 4 = 1419$ units every $T$. This represents an approximate 72-fold improvement compared to the surface code case.

This demonstrates how nonlocal connectivity enables efficient utilization of system resources, resulting in a significantly enhanced EPR generation rate.

  \section{conclusion}

 In this article, we have explored repeater-based quantum networks and provided an overview of coding theory and fault-tolerant quantum operations. We have discussed the fault-tolerant implementation of Bell measurements and teleportation, along with their combination with error correction techniques. Various code schemes designed for fault-tolerant quantum computation can be extended for fault-tolerant quantum communication, with a focus on transmission rates. Importantly, the requirement of universality for quantum communication is not necessary. We have compared different quantum codes for teleportation-based quantum communication.

Continuous-variable quantum error-correcting codes, such as those based on Gottesman--Kitaev--Preskill (GKP) states \cite{GKP01}, can be employed in quantum communication. GKP qubits enable the measurement of phase displacement errors as continuous variables while preserving the quantum information. This property facilitates decoding techniques similar to classical received vector decoding with soft information. Additionally, one can explore the concatenation of GKP codes with general CSS stabilizer codes for improved performance and error correction capabilities.

In an advanced packet-based quantum network, the transmission of quantum messages is divided into packets, similar to classical data packets in the internet, which are routed through various nodes. However,  quantum packets cannot be cloned and the packet loss problem may put inherent limitations in packet switching network.   To overcome the challenges of packet loss, one approach is to encode the quantum message using a quantum secret sharing scheme, where each packet is assigned a share~\cite{YLZ21}. This encoding scheme ensures that the arrival of quantum packets can be verified through heralding. In the event of packet loss, the remaining shares can be utilized to regenerate the missing packet. This procedure can be seen as a variant of quantum error-correcting codes, tailored specifically for packet-based quantum networks.
One may develop a systematic method for addressing packet loss by exploring different coding schemes and studying their theoretical limits.

\section{Acknowledgments}
CYL was supported by the National Science and Technology Council in Taiwan, under Grant 
111-2628-E-A49-024-MY2, 112-2119-M-A49-007, and 112-2119-M-001-007.
%111-2628-E-A49-024-MY2, 111-2119-M-A49-004, and 112-2119-M-A49-007.


% \bibliographystyle{IEEEtran}
% \bibliography{References}

\end{document}